\documentclass[twocolumn,showpacs,preprintnumbers,amsmath,amssymb,floatfix]{revtex4}
\usepackage{graphics,epsfig,subfigure}
\usepackage{color}

\begin{document}
\renewcommand{\baselinestretch}{1.3}
\newcommand\beq{\begin{equation}}
\newcommand\eeq{\end{equation}}
\newcommand\beqn{\begin{eqnarray}}
\newcommand\eeqn{\end{eqnarray}}
\newcommand\nn{\nonumber}
\newcommand\fc{\frac}
\newcommand\lt{\left}
\newcommand\rt{\right}
\newcommand\pt{\partial}

\title{Stability and (quasi-)localization of gravitational fluctuations in Eddington-Inspired Born-Infeld brane system}
\author{Qi-Ming Fu\footnote{fuqm12@lzu.edu.cn},
        Li Zhao\footnote{lizhao@lzu.edu.cn},
        Ke Yang\footnote{yangke09@lzu.edu.cn},
        Bao-Min Gu\footnote{gubm09@lzu.edu.cn},
        Yu-Xiao Liu\footnote{liuyx@lzu.edu.cn, corresponding author}}.
 \affiliation{Institute of Theoretical Physics, Lanzhou University, Lanzhou 730000,
             China}

\begin{abstract}
Stability and localization of the gravitational perturbations for a special brane system in Eddington-inspired Born-Infeld (EiBI) gravity were studied in [Phys. Rev. D 85, 124053 (2012)]. In this paper, we show that the gravitational perturbations for a general brane system are stable, the four-dimensional graviton (massless KK graviton) can be localized on the brane, and the mass spectrum of massive KK gravitons are gapless and continuous. Two models are constructed as examples. In the first model, which is a generalization of [Phys. Rev. D 85, 124053 (2012)], the brane has no inner structure and there is no gravitational resonance (quasi-localized KK gravitons). In the second one, the background scalar field is a double-kink when the parameter in the model approaches its critical value.
Correspondingly, the brane has inner structure and some gravitational resonances appear.
\end{abstract}





\maketitle

\section{Introduction}
{
It is known that Einstein's general relativity (GR) is a metric theory of gravity, and the gravitational action is given by the Einstein-Hilbert one
\beqn
S_{\text{EH}}(g)\!\!&=&\!\!\fc{1}{16\pi G_d}\int d^{d}x \sqrt{-g}
          \big({g^{MN}R_{MN}(g)}
               -2\Lambda
          \big),~~~~ \label{Einstein-HilbertAction}
\eeqn
where $G_d$ the $d$-dimensional Newtonian gravitational constant, $g=|g_{MN}|$ is the determinant of the metric of spacetime $g_{MN}$, $R_{MN}(g)$ is the Ricci tensor of that metric, and $\Lambda$ is the cosmological constant.
Several years after Einstein published his GR,
Eddington introduced an alternative proposal for the gravitational theory in 1924
\cite{Eddington1924,Schrodinger1950}, called Eddington gravity. In this theory, only the connection $\Gamma$ is the fundamental field and the action is given by
\beqn
S_{\text{Edd}}(\Gamma)&=&\fc{b}{8\pi G}\int d^{d}x
          \sqrt{-|R_{MN}(\Gamma)|}, \label{EddingtonAction}
\eeqn
where $b$ is a parameter with mass dimension $-2$, $R_{MN}(\Gamma)$ is the symmetric part of the Ricci tensor constructed solely from the connection $\Gamma$.
Eddington's theory of gravity is equivalent or dual to Einstein's GR containing only a cosmological constant, but it is incomplete because matter is not included.

Later, inspired by Eddington's theory of gravity \cite{Eddington1924,Schrodinger1950} and Born-Infeld theory of electrodynamics \cite{Born1934}, some metric and Palatini Born-Infeld theories of gravity were presented for examples in Refs. \cite{Deser,Vollick2003,Vollick2005,Vollick2006,Banados2010}.
Here, we are interesting in the Palatini theory in Ref.~\cite{Vollick2003,Vollick2005,Vollick2006,Banados2010}, which is called Eddington-inspired Born-Infeld (EiBI) gravity. This theory couples the matter fields in the conventional way.
Instead of insisting on a purely affine action, EiBI gravity is based on a Palatini-type formulation, which means that the metric and connection are
regarded as independent physical entities. The action is given by~\cite{Vollick2003,Vollick2005,Vollick2006,Banados2010}
\beqn
S(g,\Gamma,\Psi)&=&\fc{1}{8\pi G b}\int d^{d}x
          \Big[\sqrt{-|g_{MN}+bR_{MN}(\Gamma)|} \nonumber\\
            &&-\lambda\sqrt{-g}\Big] + S_{\text{M}}(g,\Psi), \label{EiBI_action}
\eeqn
}
where
$\lambda$ is a dimensionless parameter with nonvanishing value in order for the field equations to have meaning when matter fields are
absent (see Ref.~\cite{Banados2010} for the detail), $\Gamma$ is the connection field independent of the metric, and $S_{\text{M}}(g,\Psi)$ is the matter action, in which the matter fields only couple to the metric.
A key feature of EiBI theory is that it reduces to GR when matter fields are absent, but presents a different behavior from GR in the presence of matter fields.
Similar to Born-Infeld theory of electrodynamics which removes the divergence of the self-energy of a point-like charge \cite{Born1934}, EiBI theory may avoid the cosmological singularities and some undesirable features of
Einstein's theory \cite{Banados2010,Pani2011}.
Therefore, this theory has been employed to investigate the relevant cosmological, astrophysical and other issues. For examples,
the problem of dark matter and dark energy \cite{banados2008,banados2009},
the structure of compact stars \cite{Pani2011,Pani2012,Sham2012,Sham2013},
large scale structure formation \cite{du2014},
cosmological perturbations of a homogeneous spacetime
\cite{Avelino2012a,Escamilla-Rivera2012,keyang2013,Banados1311.3828,Cho1404.6081},
black holes and strong gravitational lensing \cite{Olmo1311.0815,wei2014},
observational discrimination from general relativity \cite{Sotani2014PRD}, and the generalized gravities \cite{Makarenko2014a,Makarenko2014b}.

{Recently, it was found by Odintsov, Olmo, and Rubiera-Garcia that EiBI gravity can
be naturally extended the following $f(|\hat{\Omega}|)$ theory \cite{Odintsov2014}:
\begin{eqnarray}
S_{f}(g,\Gamma,\Psi)\!\!&=&\!\!
   \fc{1}{8\pi G b}\int d^{d}x \sqrt{-g}
          \Big[f(|\hat{\Omega}|)-\lambda\Big] \nonumber\\
         &&+S_{\text{M}}(g,\Psi), \label{EiBI_action}
\end{eqnarray}
where $\hat{\Omega}$ is the representation of ${\Omega^{M}}_{N}$ in matrix notation, and ${\Omega^{M}}_{N}$ is defined as ${\Omega^{M}}_{N}\equiv g^{MP}(g_{P N}+b R_{P N}(\Gamma))={\delta^{M}}_{N}+b g^{MP}R_{P N}(\Gamma)$. It can be seen that EiBI gravity is the case of $f(|\hat{\Omega}|)=|\hat{\Omega}|^{1/2}$. The authors focused on the family of theories $f(|\hat{\Omega}|)=|\hat{\Omega}|^n$. Some interesting results were obtained. For example, the bouncing solutions persist in all the studied models (from $n=\frac{1}{3}$ up to $n=10$), and the dynamics of GR at lower curvatures for arbitrary values of $n>0$ can be recovered smoothly \cite{Odintsov2014}.}

On the other hand, in the 1920s,
in order to unify electromagnetism and Einstein's gravity,
Kaluza and Klein (KK) first introduced the idea of extra dimensions
and assumed that the electromagnetic field originates from a part  of a five-dimensional metric tensor \cite{Kaluza1921,Klein1926}.
The KK theory opens up a way to
investigate higher dimensional theory. However, it had not been drawn
enough attention until the
developments of superstring theories in the late 1970s and 1980s.  But in these theories, the
size of extra dimensions is the order of the Planck length. So detecting the extra dimensions is hopeless.
Subsequently, Akama, Rubakov, and Shaposhnikov proposed a remarkable braneworld scenario \cite{Akama1982,Rubakov1983}. In this scenario, our four-dimensional world is a braneworld or domain wall embedded in a higher-dimensional spacetime, where
the extra dimensions can be infinite and so does not need to be compacted to the Planck
scale. The idea of braneworld has attracted more and more attention after
Arkani-Hamed-Dimopoulos-Dvali (ADD) model (with finite but large extra
dimensions) \cite{Arkani-Hamed1998,Antoniadis1998} and
Randall-Sundrum (RS) model (with a finite of infinite extra dimesion)
\cite{Randall1999,Randall1999a} proposed in the late 1990s.
These two models suggest that the standard model particles are trapped on a
four-dimensional hypersurface (braneworld) by a natural mechanism, while
gravity can propagate in extra dimensions.
In RS model the brane is infinitely thin. So
more realistic thick brane models have been taken into
consideration
\cite{Goldberger1999a,Gremm2000a,Gremm2000,DeWolfe2000,Csaki2000,Gherghetta2000,Arkani-Hamed2001a,Campos2002,Kobayashi2002,Wang2002,Charmousis2003,Bazeia2004a,Liu2007,
Dzhunushaliev2009,Dzhunushaliev2010a,Liu2011}.
These ideas have appeared as the alternatives to solve some long existing problems, such as the gauge hierarchy and the cosmological constant problems
\cite{Arkani-Hamed1998,Antoniadis1998,Randall1999,Randall1999a}.
And, they may also provide us new perspectives to understand our
Universe. For a review see Refs.\cite{Rubakov2001,Csaki2004,Cheng2010}.

Generally, the configuration of a thick brane is decided by the scalar field, the gravity theory, and the ways of scalar-gravity coupling.
With a same scalar field configuration but different gravity theories, the solutions of the brane can be different, vice versa. Hence, by investigating the braneworld models, we can make a deeper understanding not only on our four-dimensional world but also on different gravity theories. There are some investigations on braneworld models in modified gravities, see Refs. \cite{zhongyi2014,chen2012,Adam} for examples.

In Ref. \cite{yangkedewenzhang2012},
a thick braneworld model in EiBI gravity with a background scalar field was investigated.
A braneworld solution was obtained and the stability of gravitational perturbations was analyzed for a special model, where the authors considered a special relation between the scalar field $\phi(y)$ and wrap factor $a(y)$: $\phi'(y)=K a^2(y)$. It was found that the gravitational zero mode is localized on the brane and hence the four-dimensional Einstein gravity can be recovered on the brane at low energy. The gravitational perturbations are stable for this special model \cite{yangkedewenzhang2012}.

In this paper, we are interested in the stability problem of gravitational perturbations for a more general EiBI braneworld model. (The recent research in general relativity can be found in Refs. \cite{salvio2009,ahmed2013}.) It will be shown that the gravitational perturbations are stable for this general model. We will construct two typical brane models as examples, one with inner brane structure and the other without. The first model is the generalization of the special model considered in Ref. \cite{yangkedewenzhang2012} and can be solved analytically. It does not support brane solution with inner brane structure and there is no gravitational resonances. The second model leads to interesting brane solution with inner brane structure, in which the scalar field has the configuration of kink, double-kink, or anti-kink. This inner brane structure results in the gravitational resonances on the EiBI brane.

The paper is organized as follows. In Sec. \ref{The_EiBI_Theory},
we construct the five-dimensional brane models in Palatini EiBI gravity, and give the domain wall
solutions for two explicit models.
In Sec. \ref{Gravity Resonances}, gravitational fluctuations are
investigated for the general flat braneworld models in EiBI gravity. Then, by using the Schr$\ddot{\text{o}}$dinger-like equation satisfied by the gravitational fluctuations,
we analyze the localization of the gravity zero mode
and the quasi-localization of the massive gravity KK modes.
Finally, the conclusion and discussion are
presented in Sec. \ref{conclusion}.

\section{The EiBI brane models and solutions}\label{The_EiBI_Theory}

Now, we construct the five-dimensional brane models in Palatini EiBI gravity~\cite{Vollick2003,Vollick2005,Vollick2006,Banados2010,Odintsov2014} with the brane generated by a real scalar field $\phi$. We mainly consider for simplicity the case of $f(|\hat{\Omega}|)=|\hat{\Omega}|^{1/2}$, which corresponds to the action of the original EiBI gravity. The action is given by
\beqn
S&=&\int d^{5}x \sqrt{-g}
          \Big[\fc{1}{\kappa b}
                 \big(|\hat{\Omega}|^{\frac{1}{2}} - \lambda \big)\Big] + S_{\text{M}}(g,\phi) ,~~~ \label{EiBI_action}
\eeqn
where $\kappa=8\pi G_5$ with $G_5$ the five-dimensional Newtonian gravitational constant.
Note that in this paper we only consider Palatini EiBI gravity, for which the scalar field $\phi$ only couples to the spacetime metric $g_{MN}$. In this case, one can show the conservation
equation $\nabla^M T_{MN}=0$ (here $\nabla$ is compatible with the metric $g_{MN}$) and so the Einstein equivalence principle is satisfied.

{In the Palatini formulation the connection $\Gamma$ and the metric $g_{MN}$  are treated as independent fields. So, the field equations for Palatini EiBI gravity theory can be derived by varying the action (\ref{EiBI_action}) with respect to the metric field $g_{MN}$ and the connection field $\Gamma^{P}_{~MN}$, respectively.} The equations of motion can be written as follows:
{
\begin{eqnarray}
|\hat{\Omega}|^{\frac{1}{2}}q^{MN} - \lambda g^{MN} &=& -b \kappa T^{MN}, \label{EOM1}\\
q_{MN}&=&g_{MN}+bR_{MN},\label{EOM2}
\end{eqnarray}}
where the energy-momentum tensor  $T^{MN}$ is defined as the standard energy-momentum tensor: $T^{MN}\equiv-\fc{2}{\sqrt{-g}}\fc{\delta L_{\text{M}}(g,\phi)}{\delta g_{MN}}$ with indices lowered by the metric $g_{MN}$. Here $q_{MN}$ is an auxiliary metric and compatible with the connection $\Gamma$, i.e., ${\Gamma^{P}_{MN}}=\fc{1}{2}q^{PL}(q_{LM,N}+q_{LN,M}-q_{MN,L})$ is the Christoffel symbol of the auxiliary metric. Then with the use of Eq. (\ref{EOM2}), the first equation of motion (\ref{EOM1}) can be rewritten as
\beq
\sqrt{-q}~q^{MN}=
\lambda\sqrt{-g}~g^{MN}-b\kappa\sqrt{-g}~T^{MN}, \label{EOM3}
\eeq
{ where we have considered $|\hat{\Omega}|^{\frac{1}{2}}=\frac{\sqrt{-q}}{\sqrt{-g}}$.}
Note that $q^{MN}$ is the inverse of $q_{MN}$: $q^{MN} q_{MP}=\delta^N_P$. { The full action is taken as the EiBI action (\ref{EiBI_action}) with the matter part given by a scalar field:
\begin{eqnarray}
S_{\text{M}}=\int d^5x\sqrt{-g}\lt[-\fc{1}{2}g^{MN}\pt_{M}\phi\pt_{N}\phi-V(\phi)\rt],\label{scalar action}
\end{eqnarray}
where $V(\phi)$ is the scalar potential.}
Then, the matter field equation is given by
\beq
\frac{1}{\sqrt{-g}}\partial_M(\sqrt{-g}\partial^M\phi)=\frac{dV}{d\phi}. \label{EOMphi}
\eeq
{A complete set of equations of the theory are consisted of Eqs. (\ref{EOM2}), (\ref{EOM3}), and (\ref{EOMphi}). The proof of their consistence can be found in appendix \ref{Appendix A}.}

In this paper, we are interested in the static flat brane with four-dimensional Lorentz invariance, for which the most general forms of the spacetime and auxiliary metrics read as \cite{Randall1999a}
\begin{subequations}\label{metric}
\beqn
ds^2\!\!&=&\!\!g_{MN}dx^M dx^N =a^2(y)\eta_{\mu\nu}dx^{\mu}dx^{\nu}+dy^2,~~~~~~ \label{RS_metric}\\
d\tilde{s}^2\!\!&=&\!\! q_{MN}dx^M dx^N=u(y)\eta_{\mu\nu}dx^{\mu}dx^{\nu} + v(y) dy^2,~~~~~~ \label{qMN}
\eeqn
\end{subequations}
where $a$, $u$, and $v$ are functions of the extra dimension coordinate $y$,
and the background scalar field $\phi$ is also a function of $y$, to be consistent with the four-dimensional Poincar$\acute{\text{e}}$ invariance of the metric (\ref{RS_metric}). Here the function $a(y)$ in the above metric is the so-called warp factor. In the famous Randall-Sundrum brane model, it is given by $a(y)=\exp(-k|y|)$, and it is just the configuration that solves the gauge hierarchy problem \cite{Randall1999a}.

By considering the spacetime metric (\ref{RS_metric}) and the auxiliary metric (\ref{qMN}),
the nonvanishing components of the Ricci tensor $R_{MN}(\Gamma)$ and the energy-momentum tensor $T^{MN}$ are given by
{
\begin{subequations}\label{RMN}
\beqn
R_{\mu\nu} &=&\fc{u u' v'-2 v(u'^2+u u'')}{4u v^2}\eta_{\mu\nu},   \\
R_{55} &=& \fc{u u' v' +v(u'^2-2 u u'')}{u^2 v},
\eeqn
\end{subequations}
}
and
\begin{subequations}\label{TMN}
\beqn
T^{\mu\nu} &=& -a^{-2}\lt[\fc{1}{2} \phi'^2+V(\phi)\rt]\eta^{\mu\nu}, \\
T^{55} &=& \fc{1}{2}\phi'^2-V(\phi).
\eeqn
\end{subequations}
Then Eqs. (\ref{EOM2}) and (\ref{EOM3}) are reduced to
\begin{subequations}\label{EOM_EP}
\beqn
&&u=a^2+b \fc{u u' v'-2 v(u'^2+u u'')}{4u v^2}, \label{EOM_EP_1}\\
&&v=1+b \fc{u u' v' +v(u'^2-2 u u'')}{u^2 v}, \label{EOM_EP_2}
\eeqn
\end{subequations}
and
\begin{subequations}\label{Aux_Metric}
\beqn
u &=& {\Xi_{+}}^{\fc{1}{3}} ~{\Xi_{-}}^{\fc{1}{3}} a^2,\\
v &=& {\Xi_{+}}^{\fc{4}{3}}~ {\Xi_{-}}^{-\fc{2}{3}},
\eeqn
\end{subequations}
respectively,
where the prime denotes the derivative with respect to the extra dimension coordinate $y$, and $ \Xi_{\pm}=\lambda+b\kappa\big(V \pm \fc{1}{2}\phi'^2\big)$.
The explicit equation of motion for the scalar field (\ref{EOMphi}) is
\beq
 4\fc{a'}{a}\phi'+\phi'' = \fc{\pt{V}}{\pt \phi}. \label{EOM_EP_3}
\eeq
{
Now we get five equations (\ref{EOM_EP})-(\ref{EOM_EP_3}) for five functions $u$, $v$, $a$, $\phi$, and $V(\phi)$.
However, by substituting the expressions of the auxiliary metric functions $u$ and $v$ in Eq. (\ref{Aux_Metric}) into Eq. (\ref{EOM_EP}), we finally obtain three equations for three functions $a$, $\phi$, and $V(\phi)$.
}


Next, we will give the EiBI-brane solutions.
The brane system is determined by the three variables $a(y)$, $\phi(y)$, and $V(\phi)$, which obey Eqs.~(\ref{EOM_EP_1}), (\ref{EOM_EP_2}), and (\ref{EOM_EP_3}).
However, the three equations are not independent because of the conservation of the energy-momentum. So the system cannot be solved uniquely. Therefore, we need to introduce some relations between these variables or the assumption of the scalar potential $V(\phi)$.
Since the differential equations (\ref{EOM_EP}) contain third-order derivative of the scalar field $\phi(y)$ and second-order derivative of the warp factor $a(y)$,
it is very difficult to solve them analytically with a given scalar potential $V(\phi)$.

In Ref. \cite{yangkedewenzhang2012},  a relation between the warp factor and the
background scalar field, $\phi'(y)=K a^{2}(y)$, was introduced and an analytic solution was found.
In what follows, we will first introduce a generalized relation, $\phi'(y)=K a^{2n}(y)$,
and give the analytic solution.
Then, we will put forward a new relation, $\phi'(y)=K_1 a^2(y)(1-K_2 a^2(y))$, and solve the equations of motion numerically.

\subsection{Model A: $\phi'(y)=K a^{2n}(y)$}

With the expectation that the scalar is a kink solution, we assume the relation $\phi'(y)=K a^{2n}(y)$ with $K$ a constant parameter and $n$ a positive integer, with which the potential can be derived from Eq.~(\ref{EOM_EP_3}) as
\beq
V(y)=\fc{n+2}{2n}K^2a^{4n}(y)+V_0, \label{Scalar_potential}
\eeq
where $V_0$ is the integral constant representing the scalar vacuum energy density. Then, substituting (\ref{Scalar_potential}) into (\ref{Aux_Metric}), we get
\begin{eqnarray}
u(y)\!\!&=&\!\!\left(\tilde\lambda\!+\!\frac{n\!+\!1}{n}b\kappa K^2a^{4n}\right)^{\frac{1}{3}}\!\left(\tilde\lambda\!+\!\frac{1}{n}b\kappa K^2a^{4n}\right)^{\frac{1}{3}} a^2,~~~~\\
v(y)\!\!&=&\!\!\left(\tilde\lambda\!+\!\frac{n\!+\!1}{n}b\kappa K^2a^{4n}\right)^{\frac{4}{3}}\!\left(\tilde\lambda\!+\!\frac{1}{n}b\kappa K^2a^{4n}\right)^{-\frac{2}{3}},
\end{eqnarray}
where $\tilde\lambda=\lambda+b\kappa{V_0}$. It can be seen that the above expressions are very complex and the solution of the warp factor is hard to find. In order to get a simple solution, we take $\tilde\lambda=0$ by fixing the scalar vacuum energy density as $V_0=-\fc{\lambda}{b\kappa}$. Then, the auxiliary metric functions are reduced to
\begin{subequations}\label{Aux_Metric_exp}
\beqn
u(y)\!&\!=\!&\! \alpha \left(\frac{n+1}{n^2}\right)^{{1}/{3}} ~a(y)^{\fc{8n}{3}+2}, \label{Aux_Metric_expa}\\
v(y)\!&\!=\!&\! \alpha \left(\frac{n+1}{\sqrt{n}}\right)^{{4}/{3}}~a(y)^{\fc{8n}{3}},\label{Aux_Metric_expb}
\eeqn
\end{subequations}
with $\alpha=(b\kappa K^2 )^{{2}/{3}}$. Now, substituting Eq.~(\ref{Aux_Metric_exp}) into Eq.~(\ref{EOM_EP}) yields
\beqn
1-\left(\frac{n+1}{n^2}\right)^{{1}/{3}}\alpha~a^{\frac{8n}{3}}(y)-\frac{b(4n+3)^2}{3(n+1)}\frac{a'^2(y)}{a^2(y)} \nonumber\\
-\frac{b(4n+3)}{3(n+1)}\frac{a{''}(y)}{a(y)}=0,\quad\quad\quad\\
1-\left(\frac{n+1}{n^2}\right)^{{1}/{3}}(n+1)\alpha~a^{\frac{8n}{3}}(y) \quad \nonumber\\
-\frac{4b}{3}(4n+3)\frac{a''(y)}{a(y)}=0,\quad\quad\quad
\eeqn
where the prime denotes the derivative with respect to $y$. The solution is
\beqn
a(y)\!&\!=\!&\!\text{sech}^{\fc{3}{4n}}\left({ky}\right),\label{Sol_Warp_factor}\\
\phi(y)\!&\!=\!&\!\frac{2K}{k}
\Big(i \text{E}(iky/2,2)
  +{\text{sech}^{\frac12}(ky)}~\text{sinh}(ky)\Big),~~~~\label{Sol_Scalar}
\eeqn
with
\beqn
K&=&\pm \frac{(1+4n/3)^{3/4}}{(n+1)} \sqrt{\frac{n}{b\kappa}} ,\\
k&=&\fc{2n}{\sqrt{3b(4n+3)}}.
\eeqn
The function $\text{E}$ in the solution (\ref{Sol_Scalar}) is an elliptic integral function. It can be seen that the scalar filed has the configuration of a kink with $\phi(\pm\infty)=\pm2 \text{Re}{(\text{E}(2))} K/k=\pm1.19814 K/k\equiv \pm v_0$. The positive $K$ corresponds to the kink solution and the negative to the anti-kink. In this paper we only consider the kink solution.
When $n=1$, the solution given in (\ref{Sol_Warp_factor}) and (\ref{Sol_Scalar}) is reduced to the one found in Ref. \cite{yangkedewenzhang2012}:
 $a(y)=\text{sech}^{3/4}(\frac{2}{\sqrt{21b}}y)$,
 $\phi(y)=\pm\frac{7^{5/4}}{2\times3^{1/4} \sqrt{\kappa}}
             (i\text{E}(\frac{iy}{\sqrt{21b}},2)
          +\text{sech}^{1/2}(\frac{2y}{\sqrt{21b}})
           ~\text{sinh}(\frac{2y}{\sqrt{21b}}))$.

For the above solution, the scalar potential (\ref{Scalar_potential}) reads as
\beq
V(y)=\fc{(n+2){(1+4n/3)^{3/2}}}{2{(n+1)^2}{b\kappa}}
     \text{sech}^3\left({ky}\right)-\fc{\lambda}{b\kappa}. \label{ScalarPotentialVy}
\eeq
From the relation $\phi'(y)=K a^{2n}$ and (\ref{Scalar_potential}), we have
$\fc{\pt{V}}{\pt \phi}=2(n+2)K a^{2n-1} a'$ and $\fc{\pt^2{V}}{\pt \phi^2}=2(n+2)\big((2n-1) a^{-2}a'+a^{-1} a''\big)$.
Then, with the expression of the warp factor (\ref{Sol_Warp_factor}), it is easy to show that $\fc{\pt{V}}{\pt \phi}=0$ and $\fc{\pt^2{V}}{\pt \phi^2}>0$ when $\phi=\pm v_0$. Therefore, $\phi(\pm\infty)=\pm v_0$ are the two vacua of the scalar potential, and the scalar field with kink configuration connects the two vacua. This is the same as the thick Randall-Sundrum brane world scenario in the frame of general relativity.

\begin{figure}[htb]
\begin{center}
\subfigure[$a(y)$]  {\label{wrapfactor1}
\includegraphics[width=4cm]{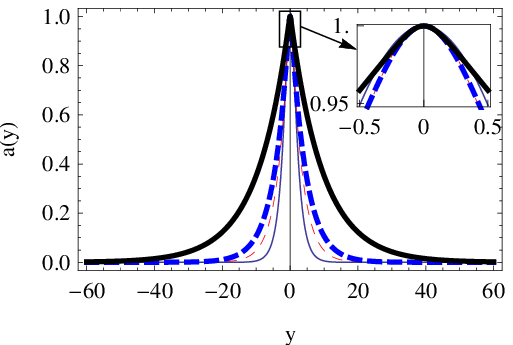}}
\subfigure[$\phi(y)$]  {\label{scalarfield1}
\includegraphics[width=4cm]{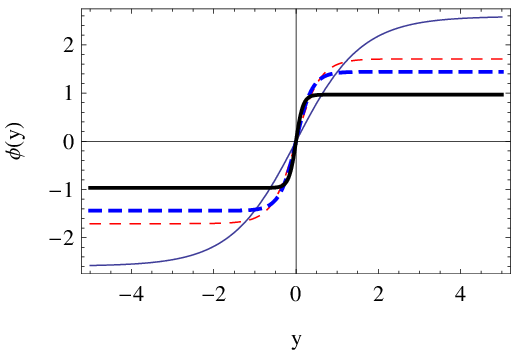}}
\subfigure[$\rho(y)$]  {\label{rho0(y)}
\includegraphics[width=4cm]{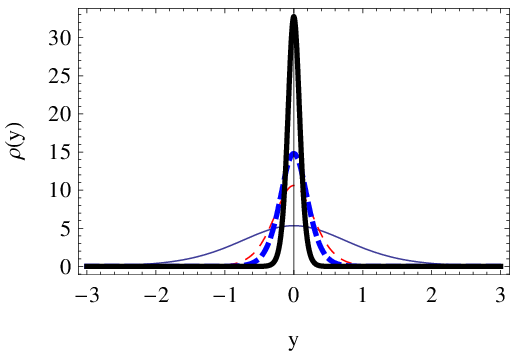}}
\subfigure[$V(y)$]  {\label{V0(y)}
\includegraphics[width=4cm]{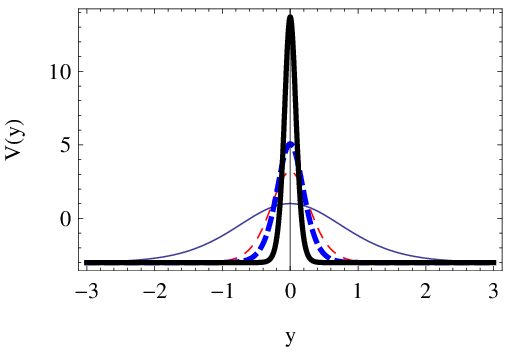}}
\end{center}
\caption{The shapes of the wrap factor $a(y)$, scalar field $\phi(y)$, energy density $\rho(y)$, and scalar potential $V(y)$ for the model A. The parameter $n$ is set to $n=1,5,10,50$, with  the thicker line corresponds to the larger $n$. The other parameters are set to $b={1}/{3}$, $\kappa=1$, and $\lambda=1$.}
\label{wrapscalar}
\end{figure}

To check if the above system describes a thick brane world scenario, we calculate the energy density of the system, which is defined as $\rho=T_{MN}U^M U^N-V_0$ with $U^M$ the velocity of the static observer. It is given by
\beq
\rho=a^2(y) T_{00}-\frac{\lambda}{b\kappa} 
    =\frac{n+1}{n}K^2 \text{sech}^3(ky),
\eeq
which indeed denotes a thick brane world located around $y=0$.
The thickness of the brane can be approximated as $d=1/k={\sqrt{3b(4n+3)}}/{(2n)}$.
For $n=1$ we have $d={\sqrt{21b}}/2$, while for large $n$ we can write $d=\sqrt{3b/n}$.
Hence, the brane becomes thinner with the increase of the parameter $n$.
The behavior of the solution with different $n$ is shown in Fig. \ref{wrapscalar}.

As is well known, in Einstein's gravity, the kink configuration of a scalar field will result in an asymptotic
AdS spacetime. Now we analyze the asymptotic structure of the five-dimensional spacetime in EiBI gravity theory considered in this section.
With the warp factor (\ref{Sol_Warp_factor}) and the functions of the auxiliary metric (\ref{Aux_Metric_exp}), the Ricci
scalar curvature reads as
\begin{eqnarray}
R&=&g^{MN}R_{MN}(\Gamma)=\frac{1}{3b(n+1)}[4n(n+2) \nonumber\\
 &-&(4n+3)(n+5)\tanh^2(ky),
\end{eqnarray}
from which, we have $R(y\rightarrow\pm\infty)\rightarrow-5/b<0$. It means that the bulk spacetime is
asymptotically AdS at the boundary of the extra dimension. This is consistent with the
brane configuration that matter mainly distributes on the brane
and AdS vacuum left far away from it.

\subsection{Model B: $\phi'(y)=K_1\, a^2(y)\big(1-K_2 \,a^2 (y)\big)$}

For the purpose of constructing a double-kink solution, we suppose
\beq
\phi'(y)=K_1\, a^2(y)\big(1-K_2 \,a^2 (y)\big), \label{ModelB}
\eeq
with $K_1$ and $K_2$ real parameters. Then, Eq.~(\ref{EOM_EP_3}) can be easily solved as
\beq
V(y)=\frac{3}{2}K_1^2a^4-\frac{7}{3}K_1^2K_2a^6+K_1^2 K_2^2 a^8 + V_0,
\eeq
where the integral constant $V_0$ represents the scalar
vacuum energy density. Thus, Eq.~(\ref{Aux_Metric}) can be
expressed as
\begin{subequations}\label{Aux_Metric_exp2}
\beqn
u(y)\!&\!=\!&\!6^{-\frac{2}{3}} \Upsilon_1^{\frac{1}{3}} \Upsilon_2^{\frac{1}{3}} a^2, \label{Aux_Metric_exp2a}\\
v(y)\!&\!=\!&\!6^{-\frac{2}{3}} \Upsilon_1^{\frac{4}{3}} \Upsilon_2^{-\frac{2}{3}}, \label{Aux_Metric_exp2b}
\eeqn
\end{subequations}
where
\begin{eqnarray}
\Upsilon_1 \!&\!=\!&\! 12bK_1^2\kappa a^4\!-\!20bK_1^2K_2\kappa a^6 \!+\!9bK_1^2K_2^2\kappa a^8\!+\!6\tilde{\lambda},~~~~ \\
\Upsilon_2 \!&\!=\!&\! 6bK_1^2\kappa a^4\!-\!8bK_1^2K_2\kappa a^6\!+\!3bK_1^2K_2^2\kappa a^8\!+\!6\tilde{\lambda},~~~~
\end{eqnarray}
and $\tilde{\lambda}=\lambda+b\kappa{V_0}$.
With the same trick, we also fix the integral constant $V_0$ by setting $\tilde\lambda=0$ to simplify the calculation, namely, $V_0=-\fc{\lambda}{b\kappa}$. Then, the auxiliary metric can be simplified as
\begin{eqnarray}
u(y)&=&(\alpha/6)^{\frac{2}{3}}
       \left(12-20K_2a^2+9K_2^2a^4\right)^{\frac{1}{3}} \nonumber\\
       &&\left(6-8K_2a^2+3K_2^2a^4\right)^{\frac{1}{3}} a^{\frac{14}{3}} ,\\
v(y)&=&(\alpha/6)^{\frac{2}{3}}
       \left(12-20K_2a^2+9K_2^2a^4\right)^{\frac{4}{3}} \nonumber\\
       &&\left(6-8K_2a^2+3K_2^2a^4\right)^{-\frac{2}{3}} a^{\frac{8}{3}},
\end{eqnarray}
where the parameter $\alpha=bK_1^2\kappa$.

Equation (\ref{EOM_EP}) can
be solved numerically with the following initial conditions:
\begin{eqnarray}
a(0)=1,\quad a'(0)=0. \label{InitialConditions}
\end{eqnarray}
We then study the behavior of the warp factor $a(y)$ and scalar
field $\phi(y)$ around $y=0$. To this end, we expand the warp
factor as $a(y)=1+py^2+\mathcal{O}(y^4)$, which satisfies the
conditions given in (\ref{InitialConditions}). Then from the
relation (\ref{ModelB}), we have $\phi'(y)=K_1(1-K_2) + 2 K_1 (1 -
2 K_2) p y^2+\mathcal{O}(y^4)$. Substituting these expanded form
into Eqs. (\ref{EOM_EP}) and solving them at the lowest order, we
get
\begin{widetext}
\begin{eqnarray}
  p&=& a''(0) \nonumber= -\frac{3 ({K_2}-1)^2 \left(9 K_2^2-20 K_2+12\right)}
     {168 - 576 K_2 + 750 K_2^2 - 440 K_2^3 + 99 K_2^4}(\leq0), \label{p} \\
  \alpha &=& \pm\frac{6 \left(11 K_2^2-24 K_2+14\right)^{3/2}}{({3 K_2^2-8 K_2+6})^{1/2} \left(9 K_2^2-20 K_2+12\right)^2}(>0),\quad\quad
  K_1 = \pm \sqrt{\frac{\alpha}{b\kappa}}.
\end{eqnarray}
\end{widetext}
The parameter $K_1$ is fixed by the equations of motion, and $K_2$ is a free parameter.
For positive and negative $b$, $\alpha$ takes the positive and negative solutions, respectively.
The two solutions of $K_1$ in fact correspond to the kink and anti-kink configurations of the scalar $\phi(y)$, and both of them describe the same one system. In this paper, we only consider the positive solution for $K_1$ without loss of generality.
So $\phi'(0)=K_1(1-K_2)$ is positive for $K_2<1$ and negative for $K_2>1$. On the other hand, note that $a''(0)$ is always negative for any $K_2\neq1$. Hence, $K_2=1$ is a critical point. When $|y|\rightarrow\infty$, the asymptotic solution for the wrap factor is
\begin{eqnarray}
  a(y)\rightarrow\text{exp}(-\frac{1}{2}\sqrt{\frac{3}{7b}}|y|). \label{ay_ModelB}
\end{eqnarray}
Here we need $b>0$ in order to have the exponentially decreasing function, which is a very important feature of Randall-Sundrum brane world model and is related to the localization of four-dimensional gravity.

\begin{figure}[htb]
\begin{center}
\subfigure[$a(y)$]  {\label{wrap factor}
\includegraphics[width=4cm]{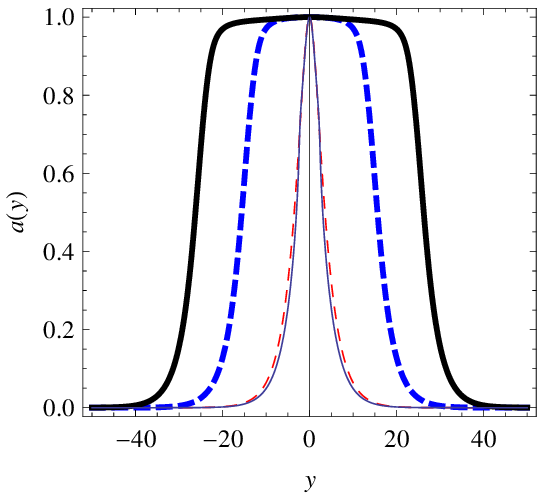}}
\subfigure[$\phi(y)$]  {\label{Scalar}
\includegraphics[width=4cm]{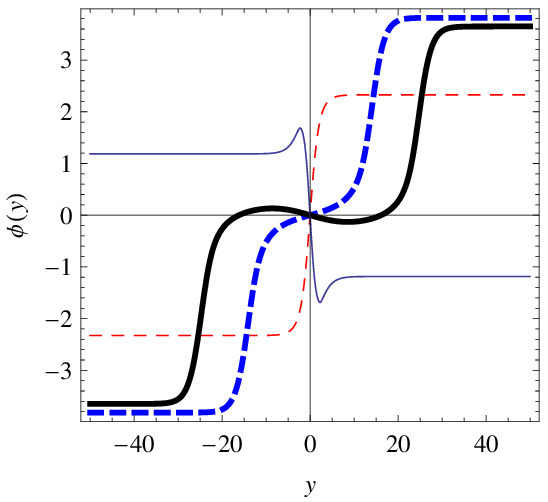}}
\subfigure[$\rho(y)$]  {\label{Fig_Energy_density}
\includegraphics[width=4cm]{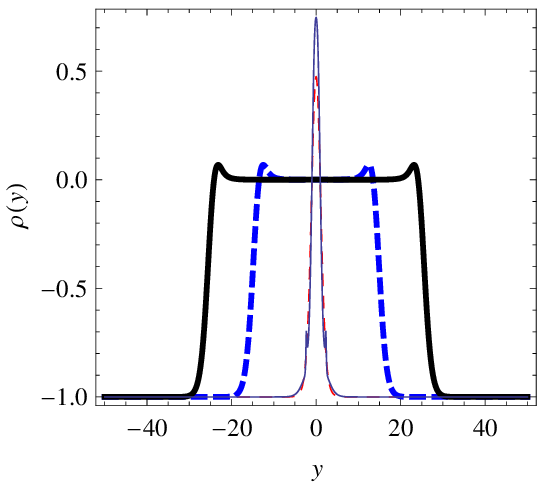}}
\subfigure[$V(y)$]  {\label{scalar potential}
\includegraphics[width=4cm]{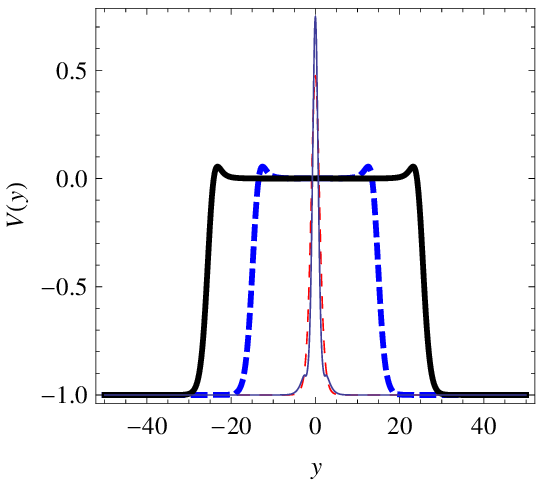}}
\end{center}
\caption{The shapes of the wrap factor $a(y)$, scalar field $\phi(y)$ , energy density $\rho(y)$, and scalar potential $V(y)$ for the model B. The
parameters are set to $b=1, \kappa=1, \lambda=1$, and $K_2=-2$ (red dashed line), $K_2=0.99$ (blue thick dashed line), $K_2=1.01$ (black thick line), $K_2=2$ (blue thin line).}
\label{A}
\end{figure}

The shapes of the solutions are shown in Fig.~\ref{A}.
It can be seen that the wrap factor and energy density become fatter first and then thinner with the increase of $K_2$, or more accurately,
 they become fatter when $K_2\rightarrow 1$. Correspondingly, the scalar field has a single kink configuration when $K_2$ far away from the first
  critical point $K_2^{(\text{c}1)}=1$ and has a double kink configuration when $K_2 \rightarrow 1$.

\begin{figure}[htb]
\begin{center}
\subfigure[$\phi(y)$]  {\label{phi(K2=1.495)}
\includegraphics[width=4cm]{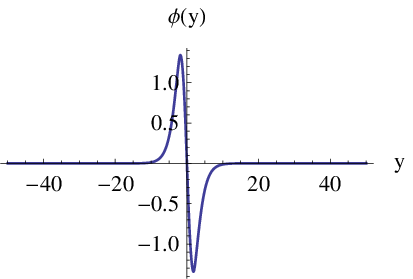}}
\subfigure[$\phi(+\infty)$]  {\label{phiinfty(K2)}
\includegraphics[width=4cm]{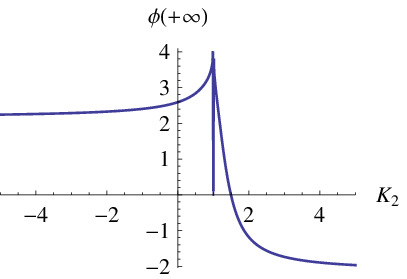}}
\subfigure[$\phi(+\infty)$]  {\label{phiinfty}
\includegraphics[width=4cm]{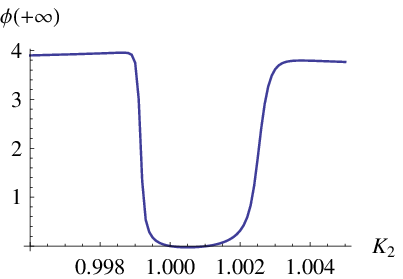}}
\end{center}
\caption{The left figure is the shape of the scalar field $\phi(y)$ with $K_2=1.495$. The second figure is about the relation between the value of $\phi(+\infty)$ and $K_2$. The third figure is $\phi(+\infty)$ around $K_2=1$. The other parameters are set to $b=1,~\kappa=1$, and $\lambda=1$.}
\end{figure}

   However, the shapes of the scalar field around the origin of the extra dimension is largely different from the case in GR,
   and they are not the standard double kink solutions.
   Note that the scalar field has an anti-kink configuration for large positive $K_2$, which can be analyzed from the numerical solution of $a(y)$
   and the expression $\phi(y)=K_1\int_0^y a^2(y)\big(1-K_2 \,a^2 (y)\big) dy$ derived from the relation (\ref{ModelB}).
   For large enough $K_2$, the integrand will be negative in some region such that the integral from 0 to $\infty$ would be negative.
   From this analysis, we can conclude that there exists the second critical point of $K_2$, called $K_2^{(\text{c}2)}$, which is larger than 1
   and decided by the parameters $\kappa$, $b$, and $\lambda$. When $K_2<K_2^{(\text{c}2)}$ we always get the kink solution, and when $K_2>K_2^{(\text{c}2)}$
    we get the anti-kink one. When $K_2=K_2^{(\text{c}2)}$, we get a critical kink solution with $\phi(0)=0$, $\phi'(0)<0$, and $\phi(\pm\infty)=0$,
     see Fig.~\ref{phi(K2=1.495)}. Besides, we also give the relation between $\phi(+\infty)$ and $K_2$ in Fig. \ref{phiinfty(K2)},
      from which we can see that $\phi(+\infty)$ increases with $K_2$ when $K_2<1-\delta_1$ or $1<K_2<1+\delta_2$, where $\delta_i$ are
      some positive constants decided by the parameters $\kappa$, $b$, and $\lambda$. However, it decreases with $K_2$ when $1-\delta_1<K_2<1$ or $K_2 > 1+\delta_2$.
      Note that there are two zero points for $\phi(+\infty)$. The first one is at $K_2=1$, for which the solution is trivial ($a(y)=1$, $\phi(y)=0$) and does not denote a brane system. The second one is at $K_2=K_2^{(\text{c}2)}$.


\section{Localization of gravity}\label{Gravity Resonances}

In a brane model, the stability of the system under the
perturbations of the spacetime metric and whether matter fields as well as
the tensor zero mode of the metric perturbations can be
localized on the brane are two important issues.
Generally speaking, the four-dimensional
massless graviton should be localized on the brane in order to
reproduce the familiar four-dimensional Newtonian potential.

The stability problem of the tensor fluctuations of the brane metric has been investigated in Ref.~\cite{yangkedewenzhang2012} only for the special model of $\phi'(y)=K a^{2}(y)$. Here, we will generalize the result of Ref.~\cite{yangkedewenzhang2012} to a general model.

Now, we consider the tensor fluctuations of the spacetime metric and auxiliary metric:
\beqn
d\hat{s}^2\!&\!=\!&\!\hat{g}_{MN}(x,y)dx^Mdx^N \nonumber\\
              &=&a^2(y)[\eta_{\mu\nu}+h_{\mu\nu}(x,y)]dx^{\mu}dx^{\nu}+dy^2,\\
d\hat{\tilde{s}}^2\!&\!=\!&\!\hat{q}_{MN}(x,y)dx^Mdx^N \nonumber\\
                      &=&u(y)[\eta_{\mu\nu}+\gamma_{\mu\nu}(x,y)]dx^{\mu}dx^{\nu}+v(y)dy^2,
\label{Fluc_Metric_UC}
\eeqn
where $h_{\mu\nu}$ and $\gamma_{\mu\nu}$ represent respectively the tensor fluctuations of the background spacetime metric and auxiliary metric, and they are transverse-traceless (TT), i.e. $\eta^{\mu\beta}\partial_{\beta}h_{\mu\nu}=0$ and $h{\equiv}\eta^{\mu\nu}h_{\mu\nu}=0$, so does $\gamma_{\mu\nu}$. For the tensor perturbation, the scalar field fluctuation is decoupled and the perturbation equation is
\beq
\fc{u}{v}h_{\mu\nu}''+\left(\frac{2u'}{v}-\frac{uv'}{2v^2}\right)h_{\mu\nu}'+\Box^{(4)}h_{\mu\nu}=0,
\label{TT_EQ_UC}
\eeq
where $\Box^{(4)}=\eta^{\mu\nu}\pt_{\mu}\pt_{\nu}$ stands for the four-dimensional D'Alembertian.
By making a coordinate transformation $dy=\sqrt{\fc{u(z)}{v(z)}}dz$, Eq. (\ref{TT_EQ_UC}) can be rewritten as
\beq \pt_{z,z}
{h}_{\mu\nu}+\fc{3\pt_z u(z)}{2u(z)}\pt_z{h}_{\mu\nu}+\Box^{(4)}{h}_{\mu\nu}=0. \label{Fluc_Eq_C}
\eeq

Further, by making the KK decomposition of ${h}_{\mu\nu}$ as
\beq
{h}_{\mu\nu}(x,z)=\varepsilon_{\mu\nu}(x)f(z)H(z)\label{KK_Decomposition},
\eeq
where
\beqn
   f(z)= \text{exp}\Big(-\int\frac{3u'}{4u}dz\Big),  \label{fz}
\eeqn
is a function which cancels the first derivative of $H(z)$ in order to form the Schr$\ddot{\text{o}}$dinger-like equation.

Inserting Eq. (\ref{KK_Decomposition}) into Eq. (\ref{Fluc_Eq_C}), we can get the following two equations
\beqn
   \Box^{(4)}{\varepsilon_{\mu\nu}(x)}
      &=& m^2 \varepsilon_{\mu\nu}(x), \label{4D_Fluctuation} \\
   \big(-\pt_{z}^2 +U(z)\big)H(z)
      &=& m^2H(z).  \label{Schrodinger_Eq}
\eeqn
Equation~(\ref{4D_Fluctuation}) is the Klein-Gordon
equation for the four-dimensional massless ($m=0$) or massive ($m\neq0$) graviton, while Eq. (\ref{Schrodinger_Eq}) is the equation of motion for the KK modes, which is a Schr$\ddot{\text{o}}$dinger-like equation with the effective potential $U(z)$ given by
\beq
U(z)=\frac{2 (\pt_{z}f(z))^2}{f^2(z)}-\frac{\pt_{z}^2f(z)}{f(z)}.\label{Effective_Potential}
\eeq
Equation~(\ref{Schrodinger_Eq}) can be rewritten as the supersymmetric form $L^{\dag} L H(z)=m^2H(z)$ with
\beqn
L&=&\Big(\frac{d}{dz}+\frac{\pt_{z}f}{f}\Big),\\
L^{\dag}&=&\Big(-\frac{d}{dz}+\frac{\pt_{z}f}{f}\Big).
\eeqn
As the operator $L^{\dag}L$ is hermitian and positive definite, this ensures that $m^2\geqslant0$ and so $m$ is real.
Thus, there is no tachyonic KK mode and it is possible to obtain a normalizable zero mode solution, which is responsible for gravity localization. By setting $m=0$ and solving Eq.~(\ref{Schrodinger_Eq}), we get the zero mode
\beq\label{zero mode}
 H_0(z)=N_0 f^{-1}(z)
 = N_0~\text{exp}\Big(\int\frac{3u'}{4u}dz\Big)
 =N_0u^{\frac{3}{4}}.
\eeq Here $N_0$ is a normalization constant. The integration of the zero mode is  expressed as
\begin{eqnarray} \int H_0^2(z)dz &=& \int H_0^2 \sqrt{\frac{v(y)}{u(y)}}dy=\int N_0^2u^{\frac{3}{2}}\sqrt{\frac{v(y)}{u(y)}}dy \nonumber\\
&=&\int N_0^2u\sqrt{v}dy< \infty.
\end{eqnarray}
Now we need to know whether the tensor zero mode is normalized on the brane for a general model. To this end, we need to analyze whether the above integration is convergent. It is not hard to see that the brane should be embedded in an asymptotic AdS spacetime, for which the asymptotic
solutions of the scalar field and scalar potential are respectively
$\phi(y\rightarrow\pm\infty)\rightarrow v_0$ and $V(y\rightarrow\pm\infty)\rightarrow V_0$,
where $v_0$ and $V_0$ are the vacuum expectation value of the scalar and a five dimensional cosmological constant. Considering
these asymptotic behaviors as well as Eq. (\ref{Aux_Metric}), we
obtain $u(y\rightarrow\pm\infty)\propto a^2(y)$ and $v(y\rightarrow\pm\infty)\propto c$ with
$c$ a positive constant. As a result,
the normalization condition reduces to $ \int
a^2(y) dy < \infty$. The asymptotic solution of the wrap factor is
$a(y\rightarrow\pm\infty)\rightarrow e^{-k_0 |y|}$ for an asymptotic
AdS spacetime, where $k_0$ is a positive  constant. Since the
integral $\int e^{-2k_0 |y|}dy$ is convergent, the tensor zero mode can be localized on the brane embedded in an AdS spacetime.

In the following two subsections, we will investigate the localization of the gravity zero
mode and the quasi-localization of the massive gravity KK modes for the brane solutions considered in previous section, respectively.

\subsection{Localization of the gravity zero mode}

For model A, from Eqs.~(\ref{Aux_Metric_expa}) and (\ref{fz}), we have
\begin{eqnarray}
f(z)=a(z)^{-\frac{4n+3}{2}}.
\end{eqnarray}
Then, the potential $U(z)$ is read as
\begin{eqnarray}
U(z)=\frac{(4n+3)\partial_{z,z}a}{2a}+\frac{(4n+3)(4n+1)\partial_z a^2}{4a^2},
\end{eqnarray}
and the Hamiltonian can be factorized as
\begin{eqnarray}
H=\Big(\frac{d}{dz}+\frac{(4n+3)\partial_z a}{2a}\Big)\Big(-\frac{d}{dz}+\frac{(4n+3)\partial_z a}{2a}\Big).
\end{eqnarray}
The zero mode is
\begin{eqnarray}
H_0(z)=N_0 a(z)^{\frac{(4n+3)}{2}},
\end{eqnarray}
where the normalization constant $N_0$  is fixed by the normalization condition $\int H_0^2(z)dz=\int H_0^2 \frac{dy}{a(z)}=\int N_0^2 a(y)^{4n+2}dy=1$. With the solution of the warp factor $a(y)=\text{sech}^{\fc{3}{4n}}\left({ky}\right)$ for model A, $N_0$ is calculated as
\begin{eqnarray}
N_0^{-2}&=&\frac{2^{\frac{3}{2 n}+4}}{2n+1}\sqrt{\frac{1}{3} b (4 n+3)} \nonumber\\
        &&{~_{2}F_1\left(\frac{3}{4} \left(\frac{1}{n}+2\right),\frac{3}{2 n}+3,\frac{3}{4 n}+\frac{5}{2},-1\right)}.~~~
\end{eqnarray}
So, the gravity zero mode is localized on the brane and the
four-dimensional gravity can be indeed recovered.

Next we consider model B. With Eq. (\ref{Aux_Metric_exp2a}), we have
\begin{eqnarray}
f(z)&=&a^{-\frac{7}{2}}(6-K_2 a^2+3 K_2^2 a^4)^{-\frac{1}{4}} \nonumber\\
     && (12-20 K_2 a^2+9 K_2^2 a^4)^{-\frac{1}{4}}.
\end{eqnarray}
Inserting the above $f(z)$ into Eq.~(\ref{Effective_Potential}), we can get the effective potential for model B:
\begin{widetext}
\begin{eqnarray}
U(z)&=&\frac{3}{4}a^{-2}
     (6 - 8 K_2 a^2 + 3 K_2^2 a^4)^{-2}
     (12 - 20 K_2 a^2 +9 K_2^2 a^4)^{-2}\times \Big[\Big(60480 - 445824 K_2
      a^2 + 1406496 K_2^2 a^4 \nonumber \\
     && - 2501856 K_2^3 a^6 + 2760996 K_2^4 a^8 ~-1944256 K_2^5 a^{10} + 856140  K_2^6 a^{12}
        - 216216 K_2^7 a^{14} + 24057 K_2^8 a^{16}\Big) (\pt_z a)^2\nonumber \\
&&~ + 2 a\Big(12096 - 77760 K_2 a^2 + 220416 K_2^2 a^4
      - 359856 K_2^3 a^6 + 370236 K_2^4 a^8  \nonumber \\
    && ~ - 245936 K_2^5 a^{10} +103080 K_2^6 a^{12} - 24948 K_2^7 a^{14}
     + 2673 K_2^8 a^{16}\Big) \pt_z^2 a\Big].
\end{eqnarray}
\end{widetext}

\begin{figure}[htb]
\begin{center}
{\includegraphics[width=6cm]{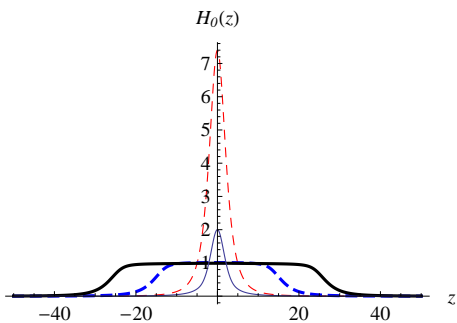}}
\caption{The gravity zero mode $H_0(z)$ for the model B.
The parameters are set to $b=1, \kappa=1, \lambda=1$, and $K_2=-2$ (red dashed line), $K_2=0.99$ (blue thick dashed line), $K_2=1.01$ (black thick line), $K_2=2$ (blue thin line).}
\label{figZeromode}
\end{center}
\end{figure}

The zero mode is
\begin{eqnarray}
H_0(z)&=&C_0 a^{\frac{7}{2}}(6-K_2 a^2+3 K_2^2 a^4)^{\frac{1}{4}} \nonumber\\
      &&(12-20 K_2 a^2+9 K_2^2 a^4)^{\frac{1}{4}},
\end{eqnarray}
where, $C_0$ is a normalization constant. For the brane solution (\ref{ay_ModelB}) obtained in the previous section, it is easy to check that the corresponding gravity zero mode for model B is normalizable: $\int H_0^2(z) dz < \infty$. Figure~\ref{figZeromode} shows the shape of the zero mode, and it shows that the zero mode is localized on the brane.

%

\subsection{Quasi-localization of the massive gravity KK modes}

\begin{figure}[htb]
\begin{center}
\subfigure[$U(z)~~(K_2=-2)$]  {\label{U(K2=-2)}
\includegraphics[width=4cm]{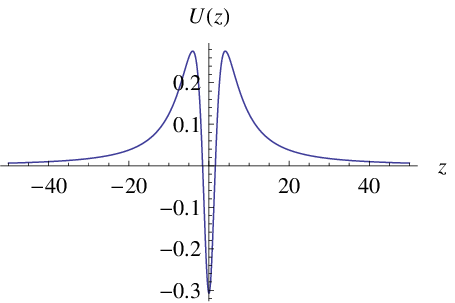}}
\subfigure[$P(m^2)~~(K_2=-2)$]  {\label{resonance(K2=-2)}
\includegraphics[width=4cm]{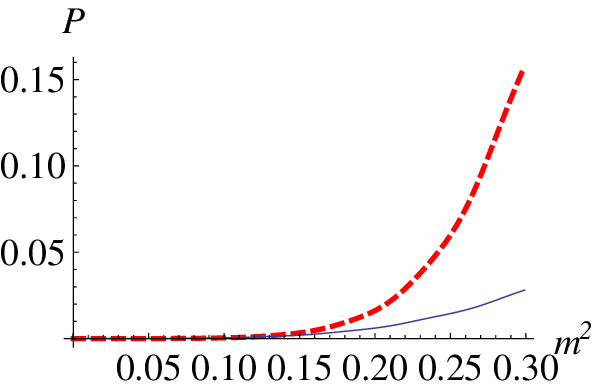}}\\
\subfigure[$U(z)~~(K_2=2)$]  {\label{U(K2=2)}
\includegraphics[width=4cm]{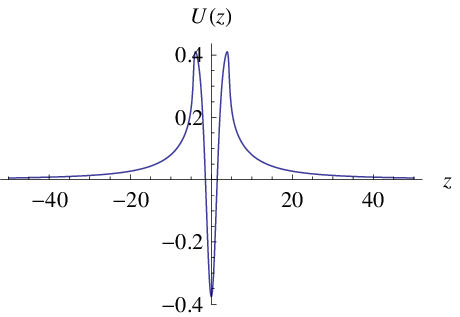}}
\subfigure[$P(m^2)~~(K_2=2)$]  {\label{resonance(K2=2)}
\includegraphics[width=4cm]{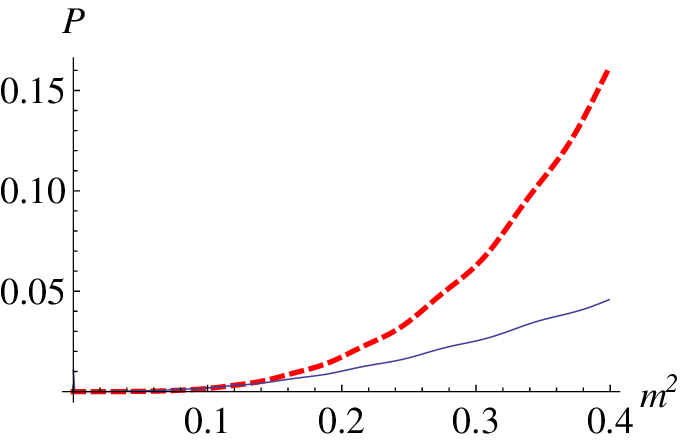}}
\end{center}
\caption{The effective potential $U(z)$ and   of the gravity KK modes for the model B.
The parameters are set to $b=1,\kappa=1$, $\lambda=1$, and $K_2=-2$ (upper) and $K_2=2$ (lower).
The red thick dashed and blue thin lines are for odd and even parities, respectively.}
\label{Gra_potential_mode1}
\end{figure}

\begin{figure}[htb]
\begin{center}
\subfigure[$U(z)~~(K_2=0.99)$]  {\label{U(K2=0.99)}
\includegraphics[width=4cm]{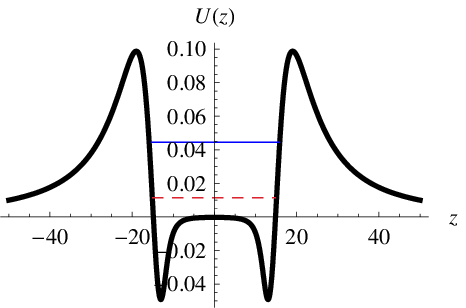}}
\subfigure[$P(m^2)~~(K_2=0.99)$]  {\label{resonance(K2=0.99)}
\includegraphics[width=4cm]{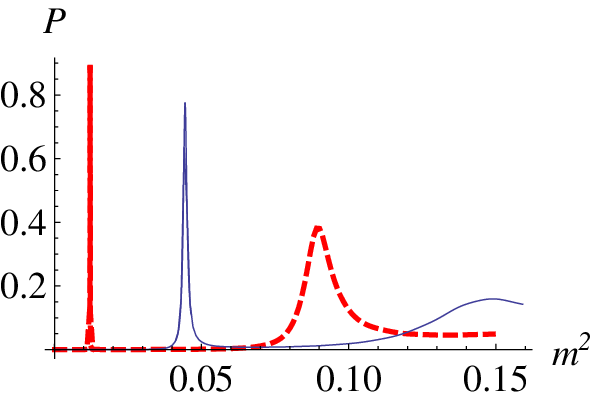}}\\
\subfigure[$U(z)~~(K_2=1.01)$]  {\label{U(K2=1.01)}
\includegraphics[width=4cm]{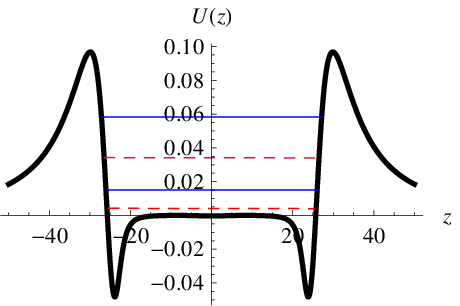}}
\subfigure[$P(m^2)~~(K_2=1.01)$]  {\label{resonance(K2=1.01)}
\includegraphics[width=4cm]{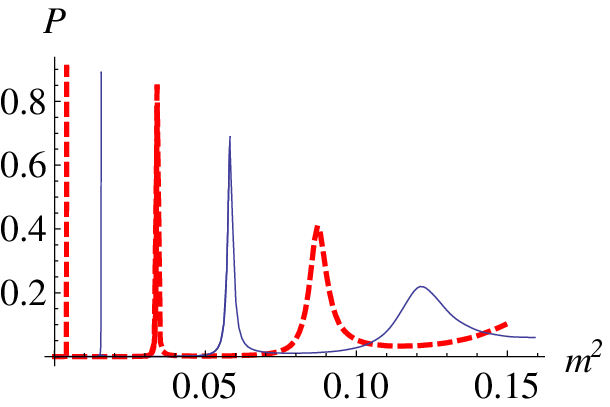}}\\
\end{center}
\caption{{The effective potential $U(z)$ and relative probability $P(m^2)$ of the gravity KK modes for the model B.
The parameters are set to $b=1,~\kappa=1,~\lambda=1$, and $K_2=0.99$ (upper) and $K_2=1.01$ (lower).
The red thick dashed and blue thin lines are for odd and even parities, respectively.}}
\label{Gra_potential_mode2}
\end{figure}

In this subsection, we mainly analyze quasi-localization of the massive gravity KK modes in model B. Figures \ref{U(K2=-2)}, \ref{U(K2=2)}, \ref{U(K2=0.99)}, and
\ref{U(K2=1.01)} show the effective potentials $U(z)$ for the KK modes of the tensor
fluctuations with $K_2=-2$, $K_2=0.99$, $K_2=1.01$, and $K_2=2$,
respectively. Analyzing the shapes of the effective potentials inspires us to investigate the possibility of
gravity resonances with the relative probability method presented in Ref.~\cite{Liu2009c} for fermions.
The gravity resonances are those massive KK modes corresponding to peaks in the relative probability $P(m^2)$ as a function of the mass square of the gravity KK modes \cite{Liu2009c}:
\begin{eqnarray}
P(m^2)=\frac{\int_{-z_b}^{z_b}|H(z)|^2dz}{\int_{-z_{max}}^{z_{max}}|H(z)|^2dz}.
\end{eqnarray}
Here $2z_b$ is approximately the coordinate width between the two maxima of the potential $U(z)$ and $z_{max}$ is set to $z_{max}$= $10z_b$. It is obvious that when $m^2\gg U_{max}$ ($U_{max}$ is the maximum value of $U(z)$), the KK mode is approximately plane wave and hence the value of $P(m^2)$ is of about $z_b/z_{max}=0.1$.
Note that, one can also use other methods used in Refs. \cite{Almeida2009}.

We investigate the massive KK modes of gravity by solving
Eq. (\ref{Schrodinger_Eq}). From Fig.~\ref{Gra_potential_mode1}, we see that
when $K_2=-2$ and $K_2=2$ there is no any peak for the massive KK modes of gravity. The further numerical calculation shows that there is no gravity resonance when $K_2$ is far away from its critical value 1.

Figure~\ref{Gra_potential_mode2} shows the effective potential
$U(z)$ and relative probability $P(m^2)$ of the gravity KK modes
with $K_2\rightarrow1$. From Fig.~\ref{resonance(K2=0.99)}, we can
see that there are two gravitational resonances for the set of
parameters $b=1,\kappa=\lambda=1, K_2=0.99$, and the mass spectra
of the resonances are calculated as
\begin{eqnarray}
m^2=\{0.012,0.044\}.
\end{eqnarray}
Figure~\ref{resonance(K2=1.01)} shows that, for another set of parameters $b=1,\kappa=\lambda=1, K_2=1.01$, there are four
resonances with the mass spectrum given by
\begin{eqnarray}
m^2=\{0.004,0.015,0.034,0.058\}.
\end{eqnarray}
Here, we only count the resonances whose mass $m$ satisfies $m^2 < U_{max}$, where $U_{max}$ is the maximum of the effective potential. The shapes of the first several KK resonances $H_n(z)~(n=1,2,3,4)$ are plotted in Figs.~\ref{P_{L,R}1} and \ref{P_{L,R}2} for $K_2=0.99$ and $K_2=1.01$, respectively. It can be seen that those modes with lower resonances mass look like bound KK modes, and they are also called quasi-bound modes. Note that we can take the $n=0$ level mode $H_0(z)$ as the four-dimensional massless graviton, which is the only one bound state.
Comparing Fig.~\ref{U(K2=0.99)} and Fig.~\ref{U(K2=1.01)}, we find that the number of the resonances increases with the
width of the potential well.

\begin{figure}[htb]
\begin{center}
\subfigure[$H_1(z)$]{
\includegraphics[angle=0,width=3.5cm]{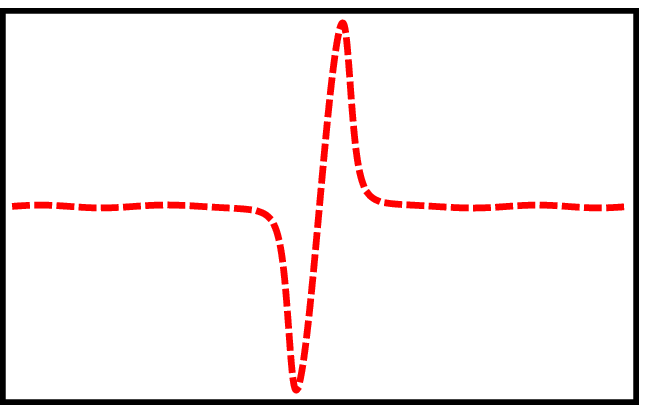}}
\subfigure[$H_2(z)$]{
\includegraphics[angle=0,width=3.5cm]{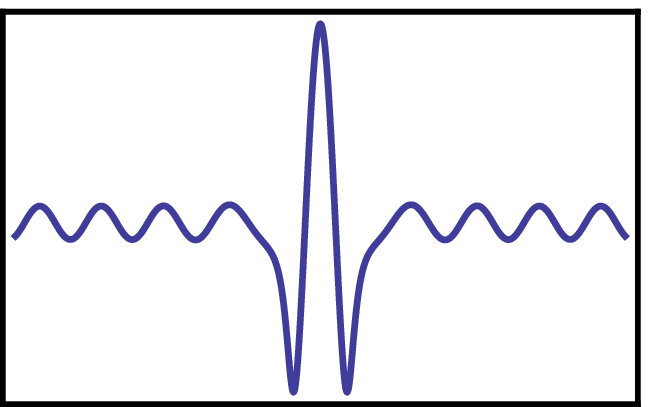}}\\
\end{center}
\caption{The lower resonances of gravity KK modes $H_1(z)$ and $H_2(z)$ for the model B. The parameter are set to $b=1,~\kappa=\lambda=1,~K_2=0.99$.}
\label{P_{L,R}1}
\end{figure}

\begin{figure}[htb]
\begin{center}
\subfigure[$H_1(z)$]{
\includegraphics[angle=0,width=3.5cm]{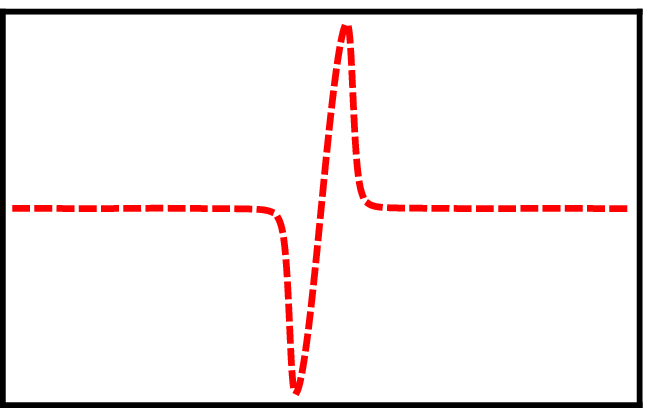}}
\subfigure[$H_2(z)$]{
\includegraphics[angle=0,width=3.5cm]{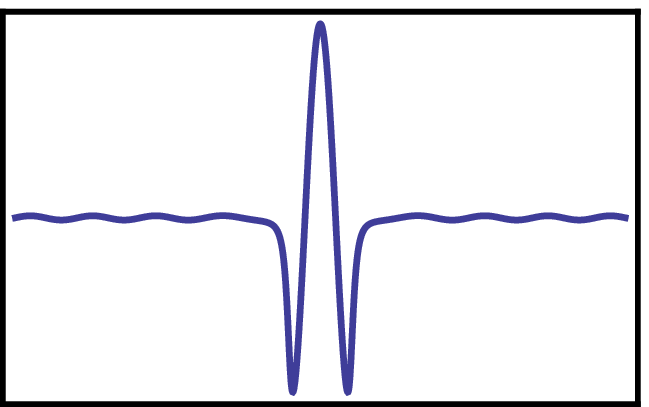}}\\
\subfigure[$H_3(z)$]{
\includegraphics[angle=0,width=3.5cm]{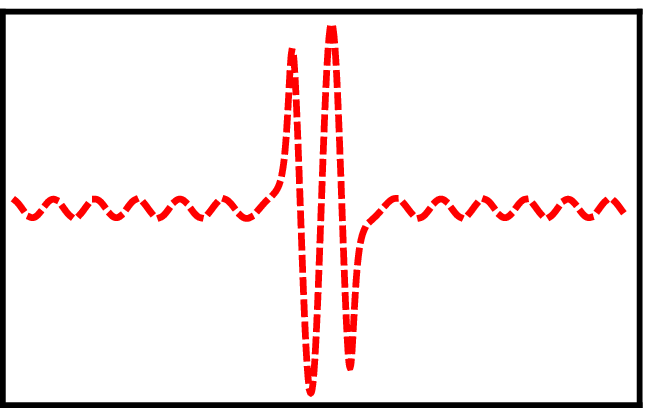}}
\subfigure[$H_4(z)$]{
\includegraphics[angle=0,width=3.5cm]{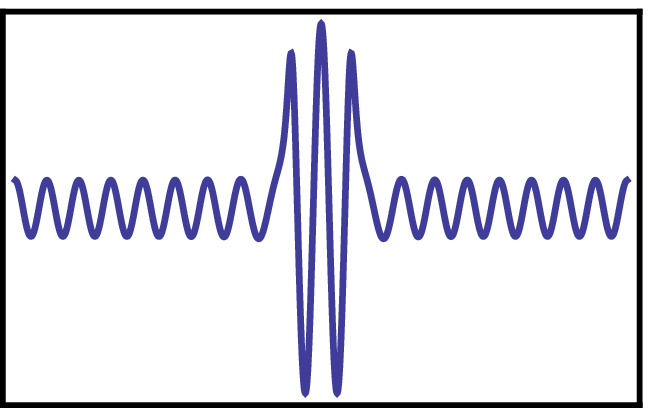}}
\end{center}
\caption{{The lower resonances of gravity KK modes $H_1(z),H_2(z),H_3(z)$, and $H_4(z)$ for the model B.
The parameter are set to $b=1,~\kappa=\lambda=1,~K_2=1.01$.}}
\label{P_{L,R}2}
\end{figure}

\section{Conclusions and Discussion}\label{conclusion}

In this paper, we investigated the localization and
resonances of gravity in the EiBI-brane system.
A general equation of motion for the gravitational fluctuations was obtained for the general bane model. This equation was converted to a Schrodinger-like equation, and the corresponding Hamiltonian could be factorized and the zero mode was also analytically solved. It was shown that the tensor perturbations are stable and the gravitational zero mode can be localized on the brane for the general brane system.

By assuming a restriction $\phi'(y)=Ka(y)^{2n}$, i.e., a generalized relation based on Ref.~\cite{yangkedewenzhang2012},
we obtained an analytical solution of the wrap factor and background scalar field. As $y\rightarrow \pm
\infty$, the scalar approaches $\pm v_{0}$, which is indeed a
kink solution with $\pm v_{0}$ corresponding to the two vacua of
the potential. The thickness of the brane decreases with the parameter $n$.
The brane is embedded in a five-dimensional AdS spacetime.

Furthermore, a relation $\phi'(y)=K_1a(y)^2(1-K_2a(y)^2)$ was investigated as another example. The parameter $K_1$ is fixed as $K_{1}=\pm
\sqrt{\frac{\alpha}{bk}}$ by the equation of motion, and $K_2$ is a dimensionless free parameter. Without loss of generality,  we considered four different cases of $K_2$:
$K_2=-2,2,0.99,1.01$. It was demonstrated that, the scalar field has a
single kink configuration which corresponds to a single brane without inner structure when $K_2$ is far away from the first
critical point $K_2^{(\text{c}1)}=1$, and has a double kink configuration which corresponds to a flat brane with inner structure when $K_2 \rightarrow
  1$. There also exists the second critical point of $K_2$, i.e., $K_2^{(\text{c}2)}$,
which is larger than 1 and decided by the parameters $\kappa$, $b$, and $\lambda$. The solution of the scalar is always a kink when $K_2<K_2^{(\text{c}2)}$, and it is an anti-kink when $K_2>K_2^{(\text{c}2)}$.

For the second model, the effective potential in the Schr\"{o}dinger-like equation for gravitational fluctuation may have an interesting inner structure: a volcano-like shape with two potential wells. Such potential structure results in a massless mode (zero mode or four-dimensional massless graviton) and a set of continuous massive modes, and may lead to
some discrete resonant KK modes. It was found that there is no resonance when the brane has no inner structure. While, the resonances will appear when $K_2\rightarrow1$, for which the brane and hence the effective potential will have inner structure. The number of the resonances increases with the width of the potential well.

The localization of fermion on the brane is an important and interesting question. In order to localize fermion on the brane, we usually need to consider the Yukawa coupling between the fermion and the background scalar field. In our models, the scalar field has kink, double kink, or anti-kink solution. So we expect that the localization of fermion on the EiBI brane can present some appealing features. This leaves for our future research.

\acknowledgments{
We would like to thank the referee for his/her helpful comments and suggestions.
We also thank Xiao-Long Du, Xiang-Nan Zhou, Feng-Wei Chen, and Bin Guo
for helpful discussions. This work was supported in part by the
National Natural Science Foundation of China (Grants No. 11075065 and No. 11375075),
 and the Fundamental Research Funds for the Central
Universities (Grant No. lzujbky-2013-18).}

{
\appendix
\section{The consistence of the system}~\label{Appendix A}

It is known that EiBI theory (\ref{EiBI_action}) can be expressed as the following bimetric action \cite{mb2009,mbpg2009,td2012}:
\begin{eqnarray}
S &=& S_G(g_{MN},q^{MN})+S_M(g_{MN},\phi) \nonumber\\
  &=& \frac{1}{2 \kappa}\int d^5 x\Big[\sqrt{-q}\Big(R(q)-\frac{3}{b}\Big) \nonumber\\
  &+&\frac{1}{b}(\sqrt{-q}q^{MN}g_{MN}-2\lambda\sqrt{-g})\Big]+S_M(g_{MN},\phi).~~~~~~ \label{BimetricAction}
\end{eqnarray}
It is obvious that the gravity action $S_G(g_{MN},q^{MN})$ in (\ref{BimetricAction}) is diffeomorphism invariant when considered in isolation, i.e., $\delta S_G=0$. We can write the variation in $S_G$ under a diffeomorphism as
\cite{carroll}
\begin{eqnarray}
\delta S_G \!=\! \int\!\! d^5 x \frac{\delta S_G}{\delta g_{MN}} \delta g_{MN}+\!\!\int\!\! d^5 x \frac{\delta S_G}{\delta q^{MN}} \delta q^{MN}.~~~~
\end{eqnarray}
Note that $\frac{\delta S_G}{\delta q^{MN}}=0$ is actually the equation of motion for $q_{MN}$ because matters do not couple with it.
If the diffeomorphism is generated by an infinitesimal vector field $V^N(x)$, the infinitesimal change in the metric is simply given by its Lie derivative along $V^N$. So we have $\delta g_{MN}=\mathcal{L}_V g_{MN}=2\nabla_{(M}V_{N)}$. Considering $\delta S_G=0$ and
$\frac{\delta S_G}{\delta q^{MN}}=0$, we have
\begin{eqnarray}
0 &=& \int d^5 x \frac{\delta S_G}{\delta g_{MN}}\nabla_M V_N \nonumber\\
  &=& -\int d^5 x V_N \nabla_M \big(\frac{\delta S_G}{\delta g_{MN}}\big),
\end{eqnarray}
where we have dropped the symmetrization of $\nabla_{(M}V_{N)}$ since $\frac{\delta S_G}{\delta g_{MN}}$ is already symmetric. Thus, we have
\begin{eqnarray}~\label{diff}
\nabla_M \big(\frac{\delta S_G}{\delta g_{MN}}\big)=\nabla_M(\sqrt{-q}q^{MN}-\lambda\sqrt{-g}g^{MN})=0.~~~~
\end{eqnarray}
Combining Eqs. (\ref{EOM2}), (\ref{EOM3}), and (\ref{diff}), we can obtain $\nabla_M T^{MN}=0$ naturally. The vanishing covariant derivative of energy-momentum tensor is consistent with the equation of motion for the scalar field. Therefore, we conclude that Eqs. (\ref{EOM2}), (\ref{EOM3}), and (\ref{EOMphi}) are consistent with each other.
}

\end{document}